Tutorial: Effective visual communication for the quantitative scientist

Marc Vandemeulebroecke, Biostatistical Sciences and Pharmacometrics, Novartis Pharma AG, Basel, Switzerland

Mark Baillie, Biostatistical Sciences and Pharmacometrics, Novartis Pharma AG, Basel, Switzerland

Alison Margolskee, Biostatistical Sciences and Pharmacometrics, Novartis Institutes for Biomedical Research, Cambridge, Massachusetts, USA

Baldur Magnusson, Biostatistical Sciences and Pharmacometrics, Novartis Pharma AG, Basel, Switzerland

All authors contributed equally to this tutorial.

Corresponding author: Marc Vandemeulebroecke, Novartis Pharma AG, Novartis Campus, 4056 Basel, Switzerland. E-mail: marc.vandemeulebroecke@novartis.com. Tel: +41 / 89 / 449 26 91.

The authors have no conflict of interest.

The authors are employees of Novartis. This tutorial has not received any funding.

Abstract

Effective visual communication is a core competency for pharmacometricians, statisticians, and more generally any quantitative scientist. It is essential in every step of a quantitative workflow, from scoping to execution and communicating results and conclusions. With this competency, we can better understand data and influence decisions towards appropriate actions. Without it, we can fool ourselves and others and pave the way to wrong conclusions and actions. The goal of this tutorial is to convey this competency. We posit three laws of effective visual communication for the quantitative scientist: have a clear purpose, show the data clearly, and make the message obvious. A concise "Cheat Sheet", available on https://graphicsprinciples.github.io, distills more granular recommendations for everyday practical use. Finally, these laws and recommendations are illustrated in four case studies.

**TUTORIAL**

# Effective visual communication for the quantitative scientist

Marc Vandemeulebroecke[1], Mark Baillie[1], Alison Margolskee[2], Baldur Magnusson[1]

**Effective visual communication is a core competency for pharmacometricians, statisticians, and more generally any quantitative scientist. It is essential in every step of a quantitative workflow, from scoping to execution and communicating results and conclusions. With this competency, we can better understand data and influence decisions towards appropriate actions. Without it, we can fool ourselves and others and pave the way to wrong conclusions and actions. The goal of this tutorial is to convey this competency. We posit three laws of effective visual communication for the quantitative scientist: have a clear purpose, show the data clearly, and make the message obvious. A concise "Cheat Sheet", available on** https://graphicsprinciples.github.io**, distills more granular recommendations for everyday practical use. Finally, these laws and recommendations are illustrated in four case studies.**

## INTRODUCTION

The goal of quantitative science is to facilitate informed decisions and actions through a data-driven understanding of complex scientific questions. It is the role of any quantitative scientist (pharmacometrician, statistician, econometrician, etc.) to support this goal through (1) appropriate quantitative methods (experimental design, statistical models, etc.) and (2) effective communication of results. There should be no conflict between technical and communicative skills – on the contrary: both of these aspects work in concert; either one without the other is not sufficient. Often, however, scientists focus on the former and neglect the latter, and sophisticated investigations remain without impact.[1] The goal of this tutorial is to help close this gap.

Scientific influence relies on effective communication,[2] and visual communication is one of the most effective channels of communication. Quoting Chambers et al,[3] "there is no single statistical tool that is as powerful as a well-chosen graph". Indeed, effective visual communication is a core competency for the quantitative scientist: he or she must not only "get the question right" (understand contextual subject matter) and "get the methods right" (technical expertise), but also "get the message right".

Visualization and the use of graphics can help at every stage of a quantitative workflow, from the very first data explorations to the final communication of conclusions and recommendations. For example, we often switch across different modes of working, from learning to confirming,[4] or from the "subjective" to the "objective" (see also Gelman and Hennig[5]). Effective visualization can help in all of these modes; see for example Gabry et al[6] for a demonstration in a Bayesian workflow. In drug development, visual communication is required at all stages from designing, analyzing and reporting clinical trials, to communicating results and supporting subsequent decision-making. The role of the quantitative scientist in this process is to ensure that relevant information (concepts, assumptions, patterns, trends, signals, conclusions) is clearly described and easy to interpret. For this, we must understand the laws and principles of effective visual communication, like the grammar of a (visual) language.[7]

If we get this right, we will be more successful in the scoping (e.g. visually clarifying the research question), execution (e.g. finding patterns in data), and communication (displaying results and conclusions) of our work. We can find important information in complex data and help people understand it. We can positively influence decisions and actions. We can create trust, partnership and engagement in cross-functional teams and audiences. We can increase our personal effectiveness. However, if we fail, we can fool ourselves and others. We can fail to see patterns in inappropriate displays. We can confuse teams, detract from the message, and pave the way to wrong conclusions and decisions.[8]

[1]Biostatistical Sciences and Pharmacometrics, Novartis Pharma AG, Basel, Switzerland; [2]Novartis Institutes for Biomedical Research, Cambridge, Massachusetts, USA

Much work has been done on this theme. Tukey,[9,10] Tufte[11] and Cleveland[12] laid the foundations for good quantitative graphics. Amit et al,[13] Bradstreet,[14] Cabanski et al,[8] Donahue,[15] Doumont,[16,17] Duke et al,[18] Few,[19] Gordon and Finch,[1] Krause and O'Connell,[20] Matange,[21] Nolan and Perrett,[22] Nussbaumer Knaflic,[23] Robbins,[24] Vandemeulebroecke et al,[25] Wainer,[26] Wong,[27] and Wong[28] have all fostered an intelligent and impactful use of visual communication and graphics. Collaborative initiatives such as CTSpedia[29] have emerged. Importantly, the theme extends beyond technical "tips and tricks" for good graphics. More fundamentally, it includes the focus on the right purpose, scientific question, situation and audience. With this in mind, it is the goal of this tutorial to distill and convey the main principles of effective visual communication for the quantitative scientist in simple, useful, and actionable terms.

The remainder of this tutorial is structured as follows. In Section 2 we posit three laws of visual communication for the quantitative scientist. Section 3 focuses on more granular recommendations for good visual display, conveniently compiled in a single page reference sheet. Complete use cases are provided in Section 4 to illustrate the application of these laws and recommendations in practice. Section 5 closes with a discussion. While most of the content is inspired by our work in pharmaceutical development, the same principles apply in any quantitative science.

## THE THREE LAWS OF VISUAL COMMUNICATION FOR THE QUANTITATIVE SCIENTIST

In quantitative sciences, effective visual communication follows three laws:

1. Have a clear purpose
2. Show the data clearly
3. Make the message obvious

These three laws correspond to the three main ingredients of any quantitative work: purpose, data, and message. Getting these right leads to success; failing in any of them leads to overall failure. In the next subsections, we discuss each of these laws in turn.

### Law 1: Have a clear purpose

*Why?*[30,31] What is the *purpose* of this display or that communication? Doumont[17] states this as the "zeroth law" of professional communication, "a principle so obvious that it had long been overlooked". Be clear and explicit about what you want to achieve. Is it to explore some data, to convey an inferential analysis, to deliver a message, convince an audience, or support a decision? It may be a mixture of these – for example, even seemingly simple exploratory plots should serve some (perhaps implicit) decision (e.g. on how to explore further). Every graph, and more generally every communication, must be tailored to its specific purpose.

It helps to carve out the scientific *question* you are trying to address, ideally in discussion with partners, and to write it down explicitly. Try not to look at any data before. This is the concept of "question-based visualizations":[25] let the scientific question determine what data to display and how. (For example, combine data from different domains if it helps address the question effectively. Do not only produce standard outputs by data domain – a display should be determined by the question it addresses, not by the way the data is organized.) As Diggle[32] put it, we "analyze problems, not data". This does not mean that the question could not be refined after seeing the data. We may well iterate over the problem space *and* the solution space – as long as we do it consciously. Senn[33] illustrates many examples of wrongly framed research questions. A common one is to focus on the wrong comparison, such as comparing a post-treatment value to the corresponding baseline value instead of to the value under a control treatment. Most quantitative graphs display comparisons,[34] and it always helps to ask "compared to what?".[11] If the comparison is not clear to the author, it will also not be clear to the reader.

Part of this first law (and of the third, see below) is also to be clear about your *audience*. Then, to adapt to your audience. Do not assume it will adapt to you. You cannot control your audience, but you can control what information and messages you deliver to it, and how. Is your audience just you (trying to see patterns in data), you in a few years (trying to remember what you did), quantitative experts such as your peers (interested in your methods), subject matter experts (eager for your main message), decision-makers (headlines only), or a mixture of these? Your visual communication will need to be different accordingly. Your communication (plot, presentation, report) is for the audience, not for you.

Clarity on the *purpose* and the scientific *question* of interest will help choose appropriate quantitative methods to address them. This, plus clarity on your *audience*, will help define the key messages and how to deliver them. (On the aspect of delivery see also Law 3 below.)

Of note, this first law is so important that it may occasionally defy other good principles. If your primary goal is to catch attention, then you may choose an iconic graphical representation that does this well, even if it violates some of the recommendations given further below.[35] However, you should never distort the data.

**Law 2: Show the data clearly**

This is Tufte's[11] maxim: "Above all else, show the data". Show it accurately and clearly. This law has several faces:

*Simplify!* "Simplify to clarify".[36] It is the prime task of quantitative scientists to make the complex simple: reveal structure in data through models, make inference through analyses, distill and convey conclusions through (visual) communication. Choose the simplest appropriate graph type; prefer familiar designs over fancy ones (see also the Cleveland-McGill effectiveness ranking in Law 3 below). Avoid fake dimensions. Make your plot "as simple as it can be, but not simpler" (attributed to Albert Einstein; also "Occam's razor" or the law of parsimony). "Understand, edit and simplify the information and design with your readers in mind".[28] Do not be confused: it is hard to make things simple. This is an iterative process: "edit and revise",[11] and repeat.

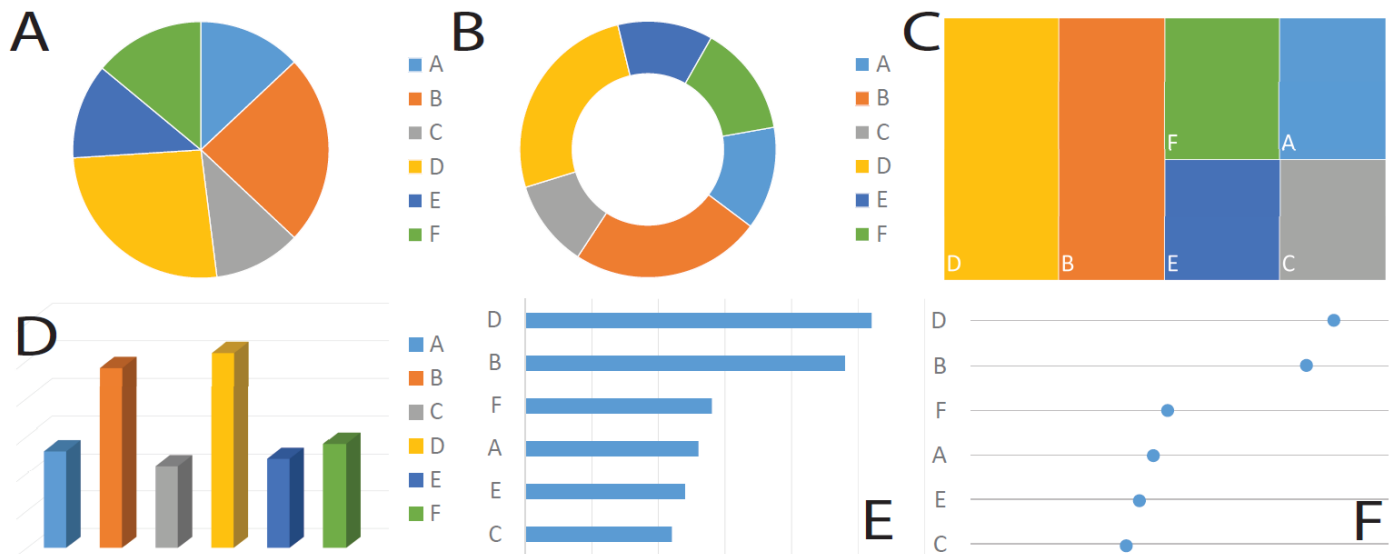

Figure 1: Law 2, show the data clearly. The pie and donut charts in panels A and B make it difficult to see the order of magnitude of some of the segments. The eye needs to compare areas, bent lengths (of the contour) or angles, graphical attributes that are not easily decoded. The donut chart even omits the angles. The mosaic plot in panel C only relies on areas; again it is hard to tell the order of magnitude. It is better to use lengths with a common baseline or positions on a common scale, such as in a barchart or dotplot (see Cleveland-McGill effectiveness ranking in Law 3). The barchart in panel D however introduces a fake dimension, which is unnecessary and makes it hard to decode the numerical values from the height of the bars. Panels E and F are appropriately simple and show the data clearly. They also order the data by magnitude to aid comparison even further. The dotplot in panel F uses minimal amount of ink and draws the eye to the position of the dots; it is the most effective way of displaying this data.

*Maximize the data-to-ink ratio* (also "data density index"[11]) within reason. Maximize the signal over the noise by removing the noise: remove anything that distracts from the purpose of the graph. Nothing is neutral: the choice of symbols or colors, background, fonts, line style, annotations. These elements are noise if they do not serve a clear purpose. Choose them wisely and parsimoniously; make the data stand out. Do not trust defaults in graphical software packages. Often, intelligent use of white space can structure a display better than a lot of ink. (The same holds for tables: these are often most effectively structured by reserving black lines for the horizontal direction and using simple alignment in the vertical direction.) Never clutter your graph with "chart junk".[37]

*Display the relevant data directly*. In a quantitative workflow, this often means to look at the raw data and not just rely on summary statistics. Cabanski[8] illustrates this with nine datasets that show completely different patterns despite identical marginal means, standard deviations and correlation coefficients (see also Anscombe[38] and Matejka and Fitzmaurice[39]). Ask yourself what is the best way of summarizing the relevant features of the data; it may not be the mean +/- standard error. When fitting a (statistical, compartmental, mechanistic etc.) model to the data to draw inference or make predictions, model-derived quantities may be the relevant data to display. In this case a plot of the raw data may be misleading if it does not account for important covariates. In a final communication, display concisely what best supports your message (see also below, Law 3).

Figure 1 illustrates some aspects of this second law. Wainer[26] has turned this law around, noting that "methods for displaying data badly have been developed for many years, and a wide variety of interesting and inventive schemes have emerged". He provides 12 highly amusing rules for "how to display data badly", with striking examples. He then concludes more seriously: "The rules for good display are quite simple. Examine the data carefully enough to know what they have to say, and then let them say it with a minimum of adornment. Do this while following reasonable regularity practices in the depiction of scale, and label clearly and fully."

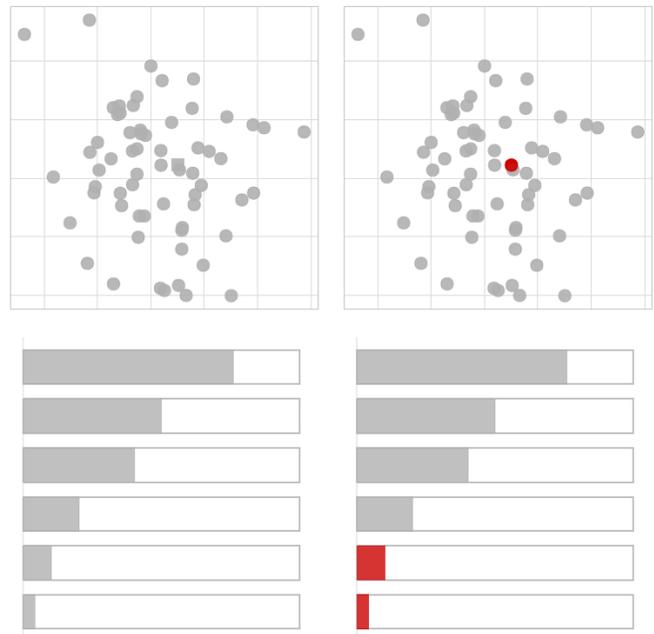

Figure 2: The power of pre-attentive processing. In the top row, one data point is marked by a different symbol (left) or color (right). Discerning the different symbol requires attention, while the different color "jumps out" pre-attentively. In the bottom row, left panel, we are drawn to a comparison of lengths. In the right panel, we introduce color to draw the eye to the bottom two bars first. While length is already a strong pre-attentive attribute, color is even stronger. We can use this to guide the viewers' attention through a plot and let them follow the story we want to convey.

**Law 3: Make the message obvious**

If the second law focused on the *data* (with a tendency to reduce noise), then the third is all about the *message* (and amplifying the signal). This assumes that you do have a message to tell, and that this message is clear at least to yourself. If there is any doubt on this, return to the first law.

The third law mandates to make your message as obvious as possible. Quoting Krzywinski and Cairo,[40] "inviting readers to draw their own conclusions is risky". Do not only make your message easy to get. Make it impossible to miss. This extends beyond graphical elements and involves all aspects of communication.

Clarity on your audience, mentioned already in the first law, is also a prerequisite for the third. It is needed for carving out the message to tell (Law 1) and for adapting its way of delivery (Law 3).

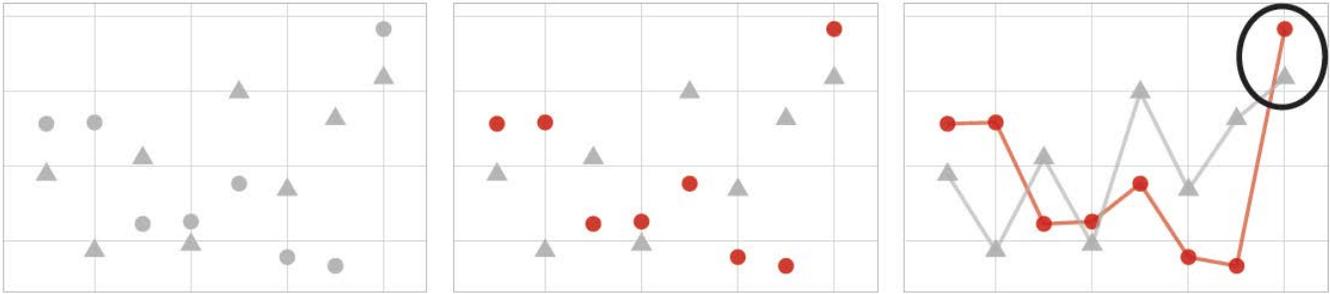

Figure 3: The principles of visual grouping: enclosure, connection, proximity and similarity. The first panel shows two groups of points, identified by similarity (of plot symbols). Here, proximity is a consequence of the data display and cannot be chosen deliberately. The second panel introduces color to further enhance the grouping by similarity (using effective redundancy and pre-attentive processing). The third panel groups the points even more effectively by connecting lines (in addition to the plot symbols and colors). Special attention is drawn to a group of points by an enclosing ellipse.

Specific examples for the third law include the following:

- Choose wisely how to encode the data you display. Color and area are good for drawing attention, but a viewer can decode positions on a common scale much more easily and accurately. Consider the effectiveness ranking of graphical attributes for encoding numerical values, as proposed by Cleveland and McGill[12,41-44] (see also Cairo,[45] Munzner,[46] and Heer and Bostock[47]). See Figure 4 for a representation of this ranking.
- Exploit pre-attentive processing as much as possible.[19, 48] Some graphical features "jump to the eye" while others require careful inspection. Consider this in your choice of how to encode the data (Cleveland-McGill effectiveness ranking, see above) and in your choice of symbols, colors, line types etc. See Figure 2 for an illustration.
- Avoid mental arithmetic. If differences or ratios are the main interest, show them directly. If both raw values and differences are of interest, consider showing both.
- Exploit the principles of visual grouping.[49] Graphical entities are most effectively grouped by enclosure, connection, proximity and similarity (in this order). That is, similar objects are perceived as belonging together, as are objects close to each other, connected by lines, or enclosed in a common subspace. See Figure 3 for an illustration: these mechanisms can provide contextual information to a plot in a simple yet powerful way.
- Minimize the viewer's eye movement. Place elements that are to be compared close to each other. Prefer direct labelling over a legend. See also Figure 5.
- Draw the reader's attention to the main points. Use appropriate graphical features (e.g. bold or colored highlighting, reference lines, circling etc.). Follow up with explicit labelling (e.g. "treatment A outperforms treatment B by X%").
- Add meaningful information to a graph to tell the whole story. E.g., include reference lines, benchmark effects, inferences etc.
- Use effective redundancy. Convey the same message through multiple channels, to amplify it and give the audience a second chance to get it. Use words and pictures in unison.[16] E.g., in addition to showing the data, consider annotating the "good" or "bad" axis direction, and state what is seen in plain words. Do not confuse redundancy (pointlessly cluttering the graph) with *effective* redundancy (conveying a message through multiple complementary channels).
- Let every plot stand on its own. Use informative labels and captions, and explain abbreviations. Do not require the reader to search through text in order to understand a figure.
- Always add a title to your plot. Phrase it as a conclusion, not a description (e.g. "plasma concentration depends on body weight" rather than "plot of plasma concentration vs. body weight").

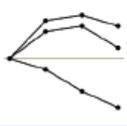

Figure 4: Selecting the right base graph; effectiveness ranking. A conscious choice of the most appropriate graph type is an essential step towards any good data display. Typical examples are illustrated here. The Cleveland-McGill effectiveness ranking of graphical attributes[12,41-44] posits that numbers are most effectively encoded by position or length, less effectively by angle or area, and least effectively by color hue or volume. Reproduced from Margolskee et al[50] with permission.

# THE GRAPHICS PRINCIPLES CHEAT SHEET

The three laws outlined above provide overarching principled advice, and should serve as a guiding star towards effective visual communication for the quantitative scientist. To further ease their implementation in practice, it helps to distill even more detailed recommendations and to illustrate them concretely.

To this end, we have introduced the Graphics Principles Cheat Sheet.[25,50] This single-page reference sheet is an integral part of this tutorial. It was carefully designed as a concise and accessible resource for everyday practical use. Yet, it draws from a wide range of sources.[7,9-12,15,16,18-20,24,26-28,41-47] We hope that it proves useful for putting the three laws into practice.

We highlight major parts of the Cheat Sheet in Figures 4-6. The full version is available on https://graphicsprinciples.github.io, along with corresponding programming code in R.[51] Figure 4 exemplifies the conscious choice of the right base graph, an essential step towards any good data display. The same figure also illustrates the Cleveland-McGill effectiveness ranking of graphical attributes (see Law 3 in Section 2). According to this ranking, numbers are most effectively encoded by position or length, less effectively by angle or area, and least effectively by color hue or volume. Figure 5 provides recommendations for facilitating comparisons and an effective use of color. Finally, Figure 6 shows some advice for displaying data more clearly and enhancing legibility and clarity of the narrative.

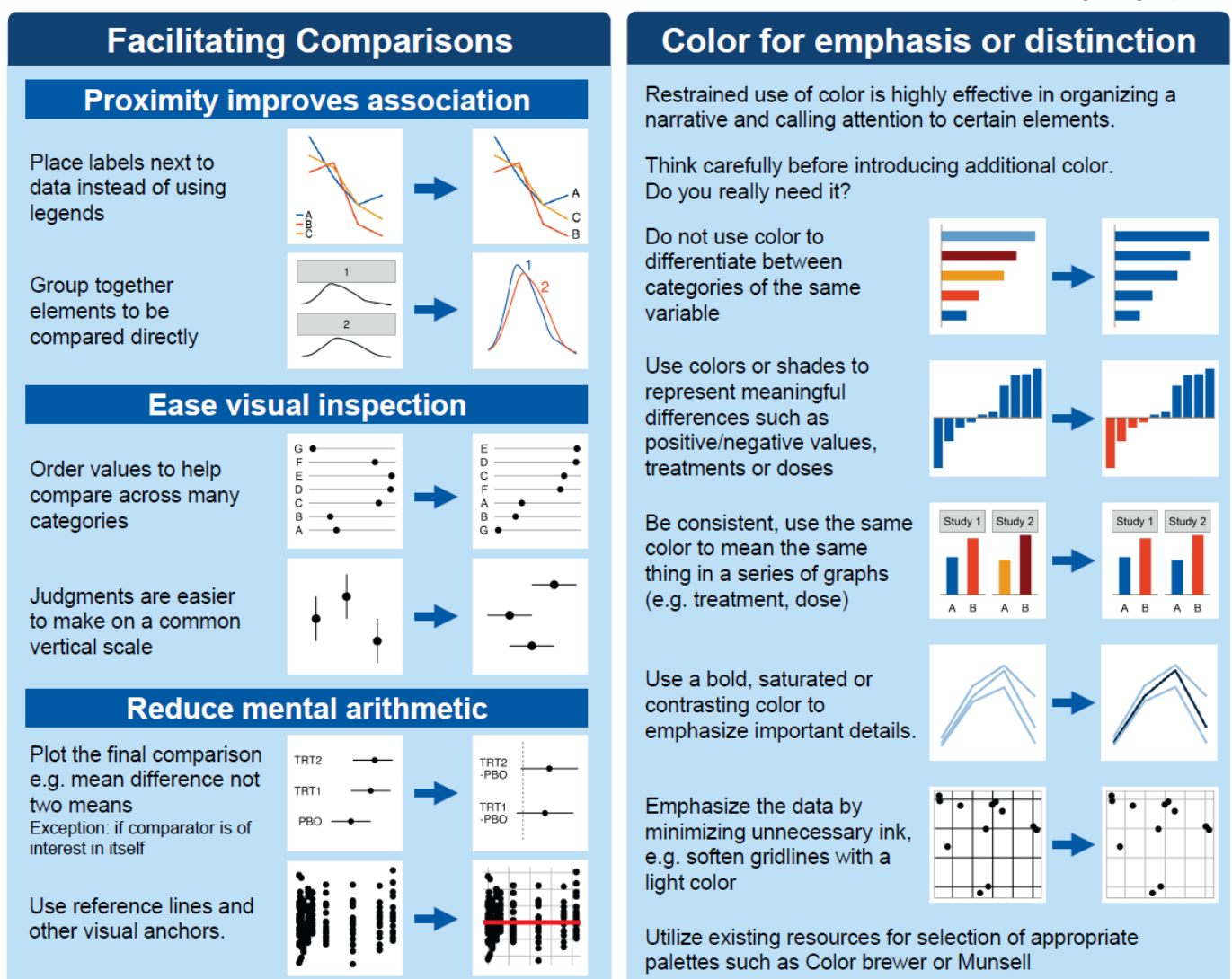

Figure 5: Facilitating comparisons; color for emphasis or distinction. Most quantitative graphs display comparisons.34 Comparisons can be facilitated by the effective use of proximity, by making visual inspections easy, and by reducing mental arithmetic. Color is a powerful stimulus. It is effective for drawing attention and organizing a narrative, but it should be used with caution and restraint. Reproduced from Margolskee et al[50] with permission.

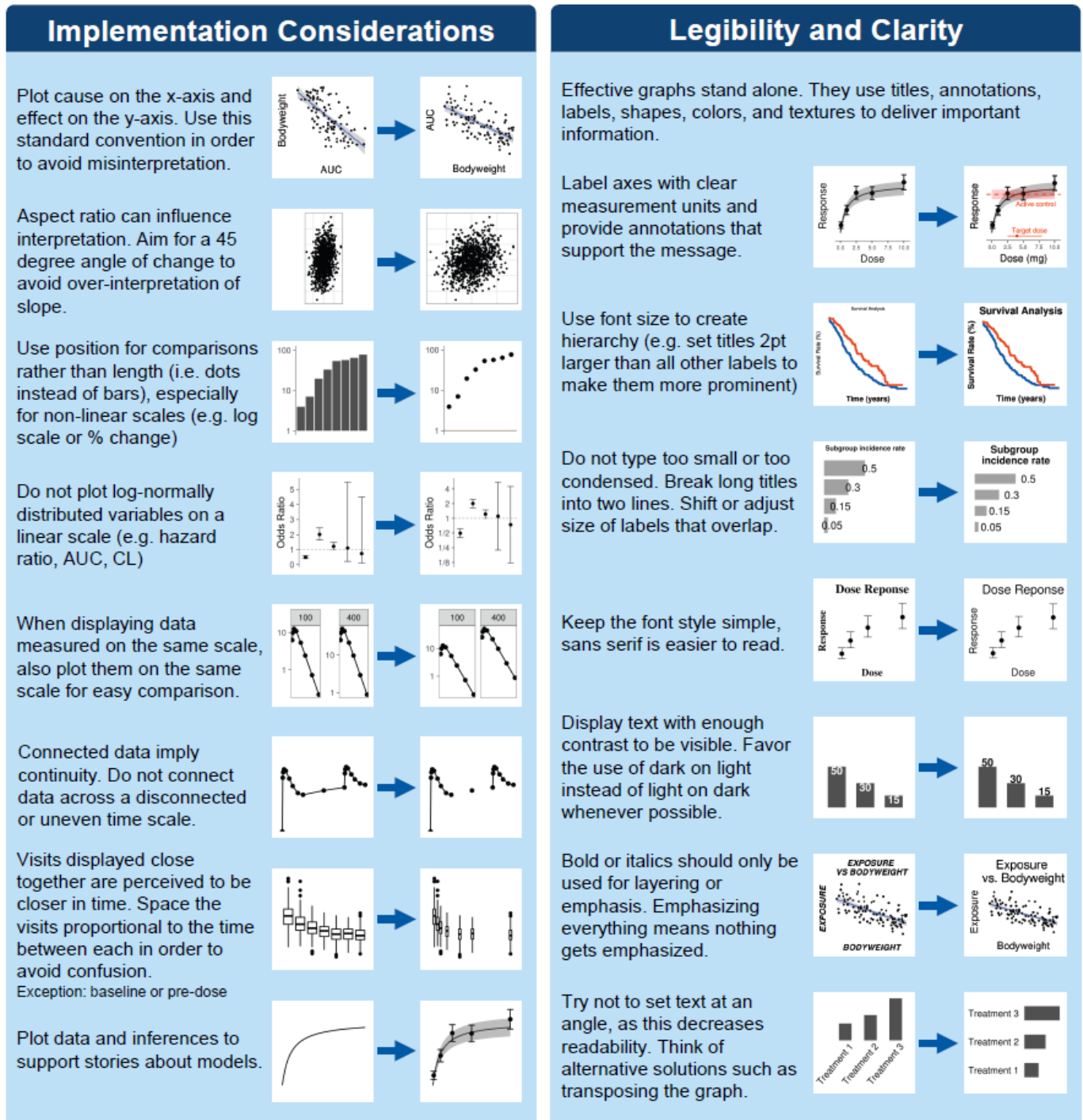

Figure 6: Implementation considerations; legibility and clarity. Various tips and tricks for displaying data more clearly. Effective graphs stand on their own; they include all necessary elements for the intended narrative. Their implementation, graphical design, and typography support legibility. Reproduced from Margolskee et al[50] with permission.

## CASE STUDIES

We now apply the laws and recommendations from the previous two sections in four short case studies, to illustrate their use in practice. The case studies are inspired by common examples from a pharmacometrician's (or medical statistician's) work. The same principles apply in any quantitative science. While the case studies resemble realistic scenarios, the data used in generating the graphs are simulated and do not represent any particular drug or trial. Details on how these case studies were generated (with programming code in R) are included in the online appendix.

### Case study 1: Exploratory exposure-response analysis

This case study illustrates the importance of understanding the scientific context when exploring data graphically. An exploratory data analysis is more than just "plotting data"; it can lead to a deeper understanding and inform next steps.[6,52] However, like an analysis that is poorly thought through, a poorly implemented graph can also deceive.

Consider an inhaled drug intended to improve lung function, with the target site of action in the lung. The drug is also absorbed systemically from the lung. Suppose that the team wants to fine-tune the choice of a recommended dose. A typical starting point for this question is often a plot of the response variable of interest against a summary measure of plasma concentration (e.g. the area under the concentration time curve, AUC). Figure 7A shows such a plot, generated using the default settings of the R package ggplot2.[53]

In terms of good graphical principles, this plot leaves a fair bit to be desired. Several improvements are warranted, including proper axis scaling, gridlines, annotation, font size, etc. One particularly egregious issue is the lack of care in selecting axis labels, leaving programming labels for the plotted variables (presumably only then to make the effort of explaining them in a caption). An improved version is shown in Figure 7B, addressing many of these formatting issues. With an added LOESS smoother,[54] we see a positive non-linear trend, suggesting a shallow sigmoidal exposure-response relationship.

It is tempting, especially when presented with a suboptimal graph, to immediately set about fixing the various graphical imperfections and produce a more appropriate and visually appealing version of the *same* graph. This is an example of selective attention,[55] focusing on the detail but overlooking the higher purpose of the task (i.e. the "why"). Instead, let us now take a step back and revisit this example in the context of the first law of visual communication: have a clear purpose.

Why are we conducting an exposure-response analysis? Recall that the scientific interest is to fine-tune the *dose*, and that the drug is inhaled and acting locally in the lung. The implicit assumption of an exposure-response analysis is one of causality. Here, however, plasma concentration is unlikely to be on the causal path from dose to response. What would be a better way to address the scientific question of interest?

Consider Figure 7C, where instead of estimating an overall trend we now look at the trends within dose. Clearly, any apparent trends within dose do not follow a consistent pattern across doses. The only reason why exposure and response appeared associated in the previous two plots is that they share a common cause, namely dose. In other words, dose is a confounder in those plots, and indeed dose is a better predictor of response than systemic concentration. We should build a dose-response model, rather than an exposure-response model, and choose a recommended dose based on this (and any information on safety and tolerability).

### Case study 2: Pharmacokinetic exposure by ethnicity

This case study is inspired by a publication comparing the pharmacokinetic exposure of a drug in Caucasian vs. Japanese subjects. The same single oral dose was administered to all subjects. The original graph displays mean +/- standard deviations (SD) of the plasma concentration over time, grouped by ethnicity (Caucasian or Japanese), as in Figure 8A.

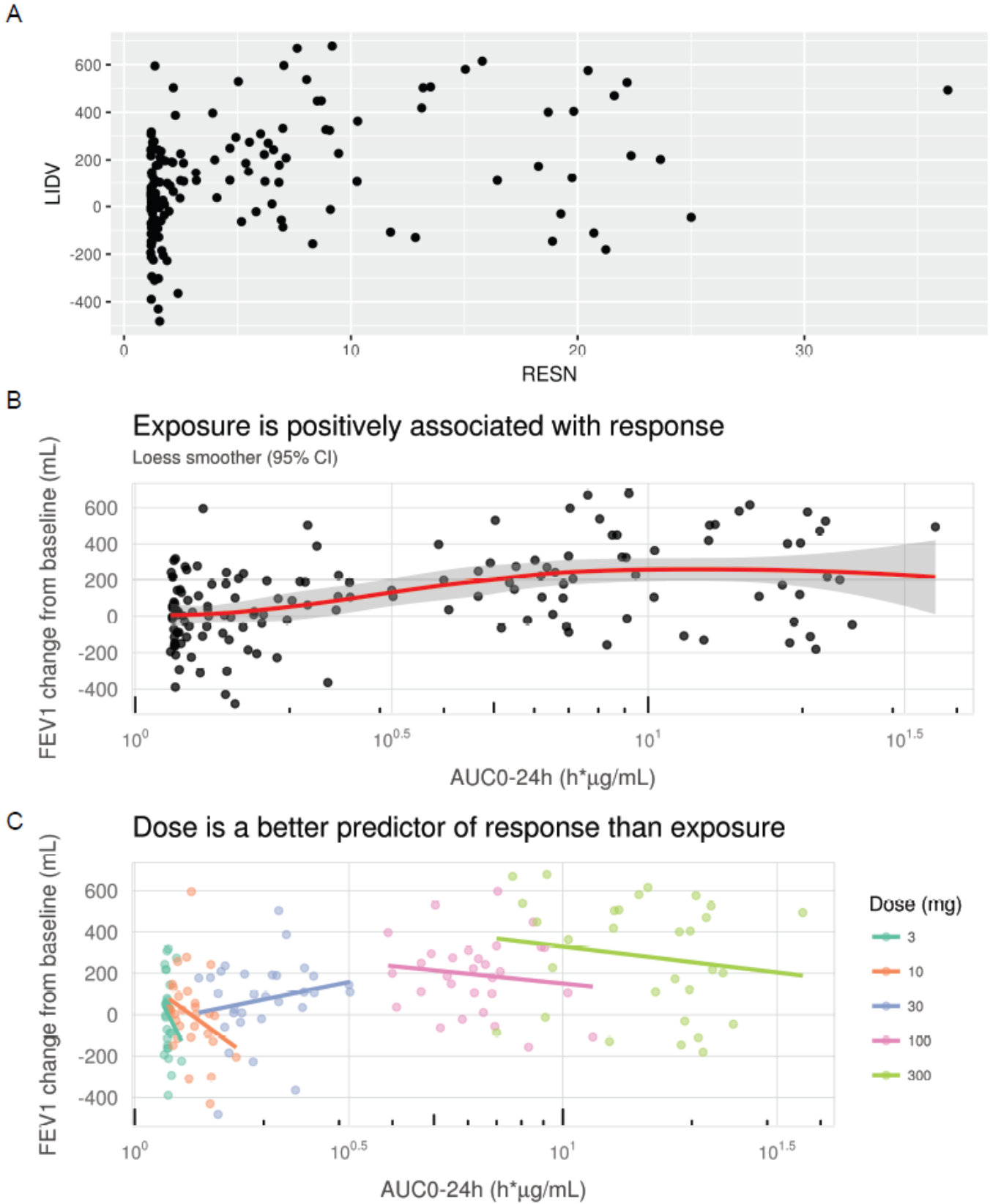

Figure 7: Exploratory exposure-response analysis. A scatterplot of response vs. exposure (A) is improved by applying good graphical principles (B), and fundamentally changed by revisiting the question of interest (C).

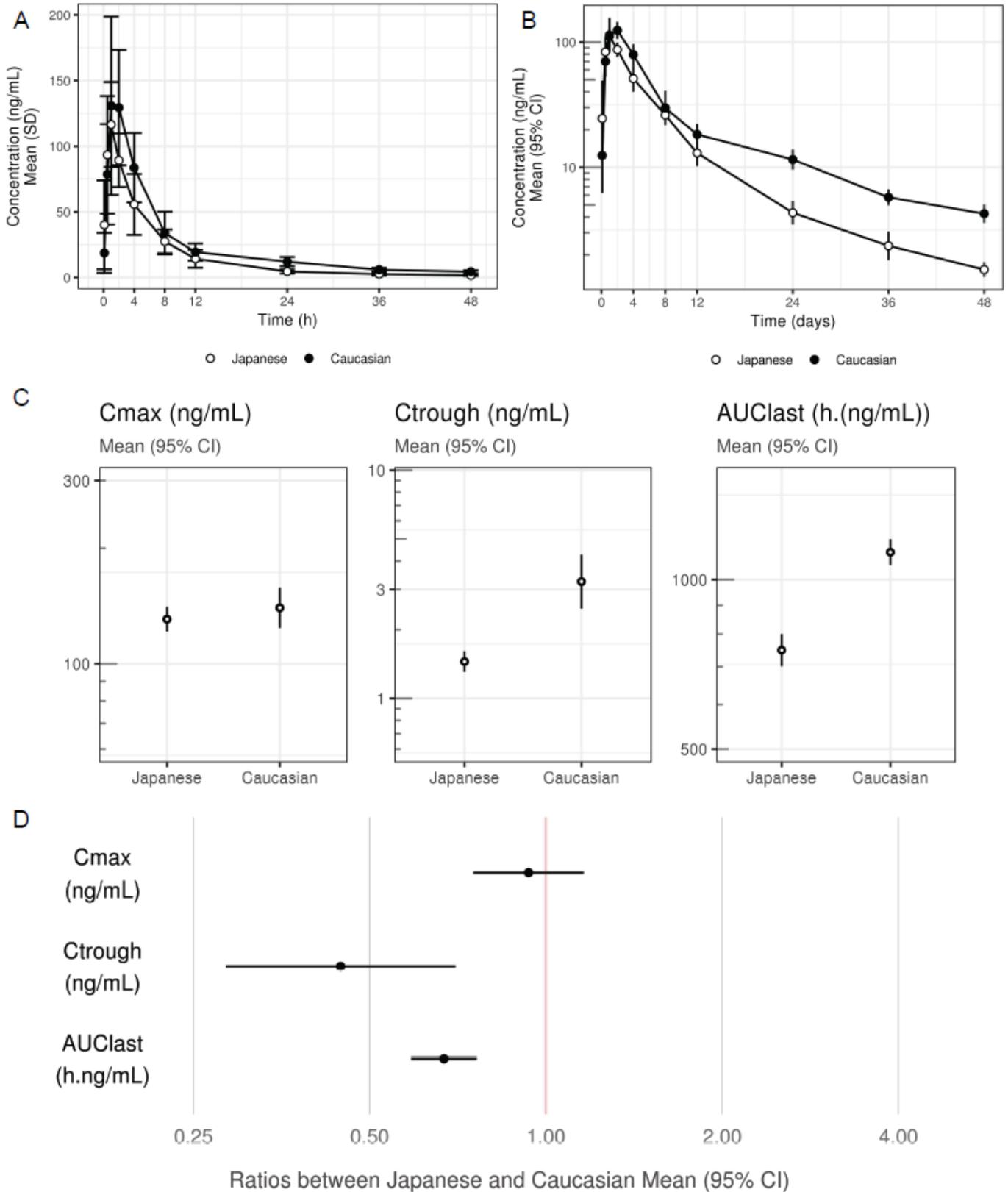

Figure 8: Pharmacokinetic exposure by ethnicity. A graph that looks fair at first sight (A) reveals important information after two simple changes, namely, scaling the y-axis differently and plotting confidence intervals instead of standard deviations (B). Key messages are made more obvious by directly plotting the pharmacokinetic parameters of interest (C), or even only their ratios (D).

This time, let us start with Law 1. What is the purpose of this graph? For drugs that are mainly developed in a Caucasian population, Japanese drug regulation requires sponsors to investigate whether the pharmacokinetics (PK) are similar or different between Caucasian and Japanese populations. The purpose of this graph is to help address this question. Looking at it, we may be tempted to say that the PK are reasonably similar. But are they really? If they are not, then in what way?

This leads us to Law 2: show the data clearly. The graphical attributes in Figure 8A appear to be wisely chosen: the symbols and labels are clear, the gridlines are supportive and stay in the background, and there is no unnecessary adornment. However, at least two things obscure the answer to our question of interest. First, the concentrations are plotted on a linear scale, which makes it difficult to distinguish them at the lower end of the profile. Concentrations should be plotted on a logarithmic scale because they are log-normally distributed. In fact, for concentrations in particular (but not generally for any log-normally distributed data) we should produce both types of display: one on a log-linear scale (to assess the elimination phase) *and* one on a linear scale (to see the peak more clearly). Second, it is hard to determine whether any differences are significant when standard deviations are plotted instead of standard errors or confidence intervals. Standard deviations show the variation in the data; they do not shrink when more data is collected. Standard errors show the variation in the means. Confidence intervals may be the best choice as they directly show the uncertainty about the means.

These issues have been fixed in Figure 8B (we only show the log-linear version). To reduce cluttering, the ticks at the end of the whiskers have also been omitted (non-data ink). If the graph displayed more than two profiles, we might consider replacing the whiskers by (shaded) confidence bands, or separating the graph out in panels or "small multiples" (see bottom of the backside of the Cheat Sheet on https://graphicsprinciples.github.io). From Figure 8B it appears that the higher concentrations are not meaningfully different, but the elimination phase does differ between the two ethnicities. This could also translate to different average exposures. That is, based on two simple changes in the plot, we now see answers emerging for our initial question about PK differences. We see them emerging with respect to three key PK characteristics: peak, elimination/trough, and overall exposure. While (depending on the drug) similarity in the peak may be reassuring from a safety point of view, a lower overall exposure in Japanese subjects could be a concern for efficacy.

Moving on to Law 3, let us now make the message obvious. We could choose a completely different graph type to hone in on the message. Figure 8C shows the three (non-compartmentally derived) quantities Cmax, Ctrough and AUClast with 95% confidence intervals by ethnicity. Clearly, Ctrough and AUC are different between the two ethnicities, but Cmax is not. We could derive these quantities from a compartmental model fit and produce the same plot. Or we could go one step further and show directly their geometric mean ratio, Japanese vs. Caucasian subjects, as in Figure 8D. This last plot answers the initial question most succinctly, and its graphical appearance has also been further simplified (no frame, minimal gridlines, mildly highlighted line of equality), to not distract from the message. In practice, Figures 8B, 8C and/or 8D together may be most informative, covering the time course as well as differences in key parameters.

**Case study 3: Improving a "waterfall plot"**

This case study illustrates the importance of aligning a graph with the scientific question it should address, the option of filtering signals through a model, and finally the display of a scientific answer in a condensed messaging graph.

Consider a small early development trial, randomized and placebo-controlled (2:1 randomization), with a continuous primary endpoint measured at baseline and longitudinally over a period of 4 weeks. Lower outcome values are better, and there are no dropouts and no missing data. Suppose that the team is interested in the effect of the drug at the last measurement time point, as it is often the case. A common approach in early development trials is to simply plot the observed change scores in a so-called "waterfall plot" such as Figure 9A.

To probe Law 1, what is the question addressed by this plot? It asks about the treatment effect *after* 4 weeks of treatment. Is this the right question? Let us assume for a moment that it is. Then a waterfall plot is not ideal for addressing it.[56] Small treatment effects are difficult to discern, especially with an unbalanced randomization ratio. The audience must observe the distribution of color across the entire plot just to determine which treatment is more beneficial; this can become even more difficult with a larger sample size or more than two treatment groups. In Figure 9A, one might see a treatment benefit, but how large is it and how certain of it are we? The popularity of waterfall plots is a mystery.

If we insist on week 4 as the only time point of interest, we could present overlaid density plots or side-by-side boxplots for a better appreciation of the difference in distribution between the two treatment arms. Figure 9B shows an example with the raw data points included, which is a much better alternative to Figure 9A. The side-by-side placement facilitates the treatment comparison, and the plot is simple, familiar and uses minimal ink for what it shows. Graphical attributes (colors, font size, etc.) are easily readable.

However, with such rich longitudinal data, it may be more informative to ask the question about the treatment effect *during* – not just *after* – the first 4 weeks of treatment. This is especially relevant in the early, more exploratory development phase (and it would be even more relevant if there were dropouts). As a rule, the recommended first step is to *visualize the totality of the data*. Figure 9C does this and includes means by treatment and time point. We see large inter-individual variability and overlap between the treatment groups. We also start to get an appreciation for the time-course of a mean effect. We see linear trajectories of the means over time, with the active arm appearing to improve and the placebo arm staying fairly constant. We cannot exclude that the apparent gap might continue to increase beyond 4 weeks of treatment. This plot, while doing little more than displaying the raw data, is already worth sharing with the project team. It facilitates a much richer understanding of the data than the previous two plots. It shows the data clearly (Law 2).

Depending on the goal of the analysis, we could stop here. But if we want to quantify the treatment difference while adjusting for important covariates, we should proceed with a statistical model. Based on Figure 9C a linear model appears appropriate. We fit a linear model with treatment, patient-specific intercept and slope, and we now also adjust for the baseline value of the primary endpoint and for any other important covariates. We can then visualize the data filtered through this model, omitting the raw data but displaying longitudinal point estimates and some form of uncertainty intervals for both treatment groups (Figure 9D). This gets closer to the nature of a messaging graph, focusing directly on the results of our model. Optionally – and depending on the audience! – we could even go one final step further and display the treatment difference directly, as in Figure 9E. This plot addresses the question about the treatment effect over time without requiring any mental arithmetic. We can read off approximate estimates for the treatment effect, and the level of confidence is easily appreciable from the added confidence band (which does include zero!). Appropriate and parsimonious annotations make the message even more obvious (Law 3), also through "effectively redundant" information (stating what can be seen).

It is worth emphasizing that this last plot should not be the only one generated, and probably not the only one shown either. Strongly reduced messaging graphs require a robust understanding of the underlying data, which can only be built through a workflow such as the one described above. Further, depending on the situation and the audience, they might be challenged as loaded or unscientific. (E.g., the apparently perfect linear trend in Figure 9E appears "unrealistic".) It is therefore important to ensure *and* emphasize that this last plot derives from a model which (as every model) is intended to separate the signal from the noise, and that the choice of this model is justified by a thorough inspection of the data.

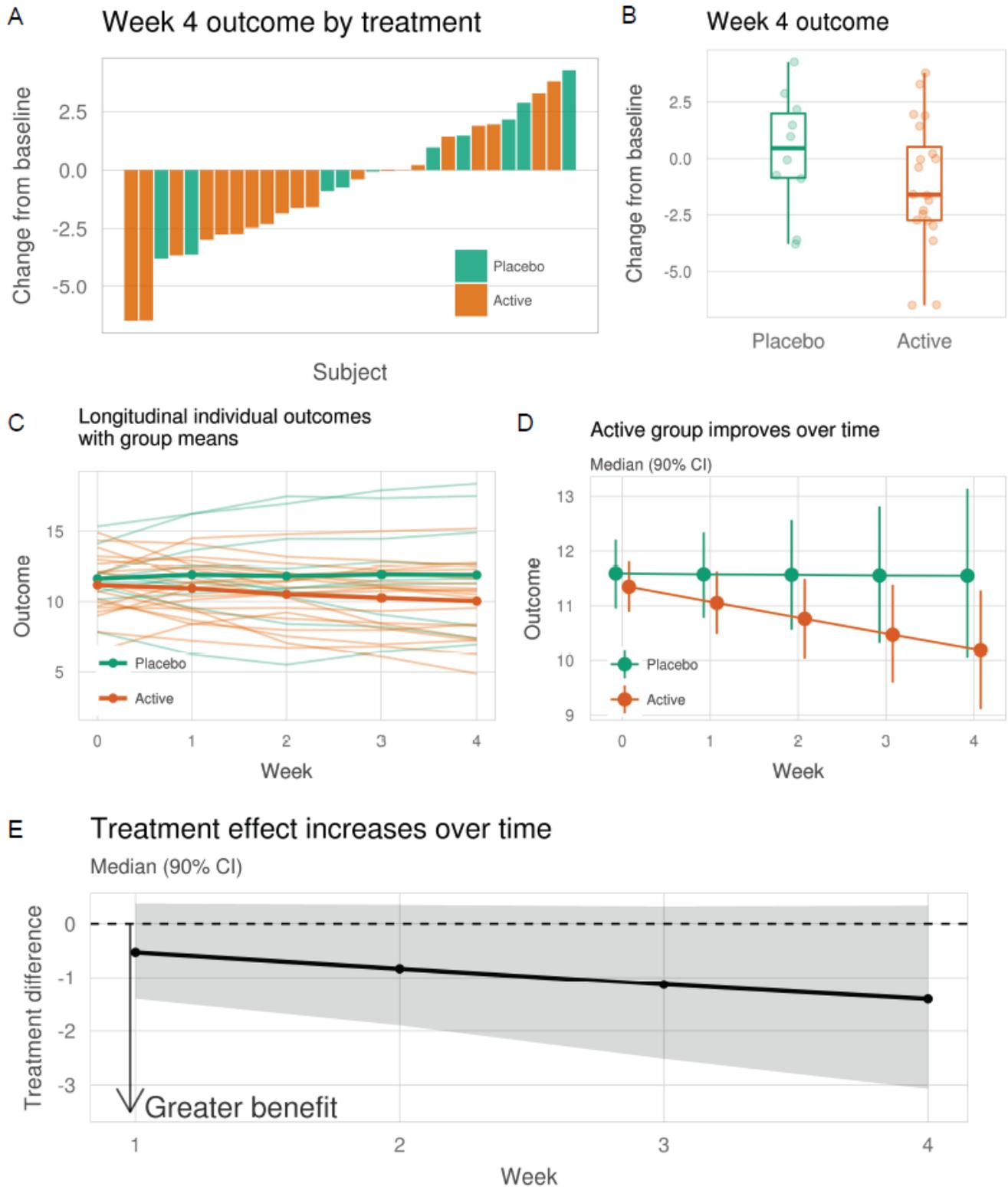

Figure 9: Improving a "waterfall plot". A waterfall plot focusing on the last observed time point (A), common in early development, is improved by a side-by-side boxplot (B) which shows the treatment comparison more clearly. Showing the totality of the data during the first 4 weeks (C) facilitates an even richer understanding, including a suggested linear trend of the treatment effect over time. This may justify the fit of a linear model (D) and ultimately (optionally) a condensed messaging graph on the treatment effect (E) based on this fit.

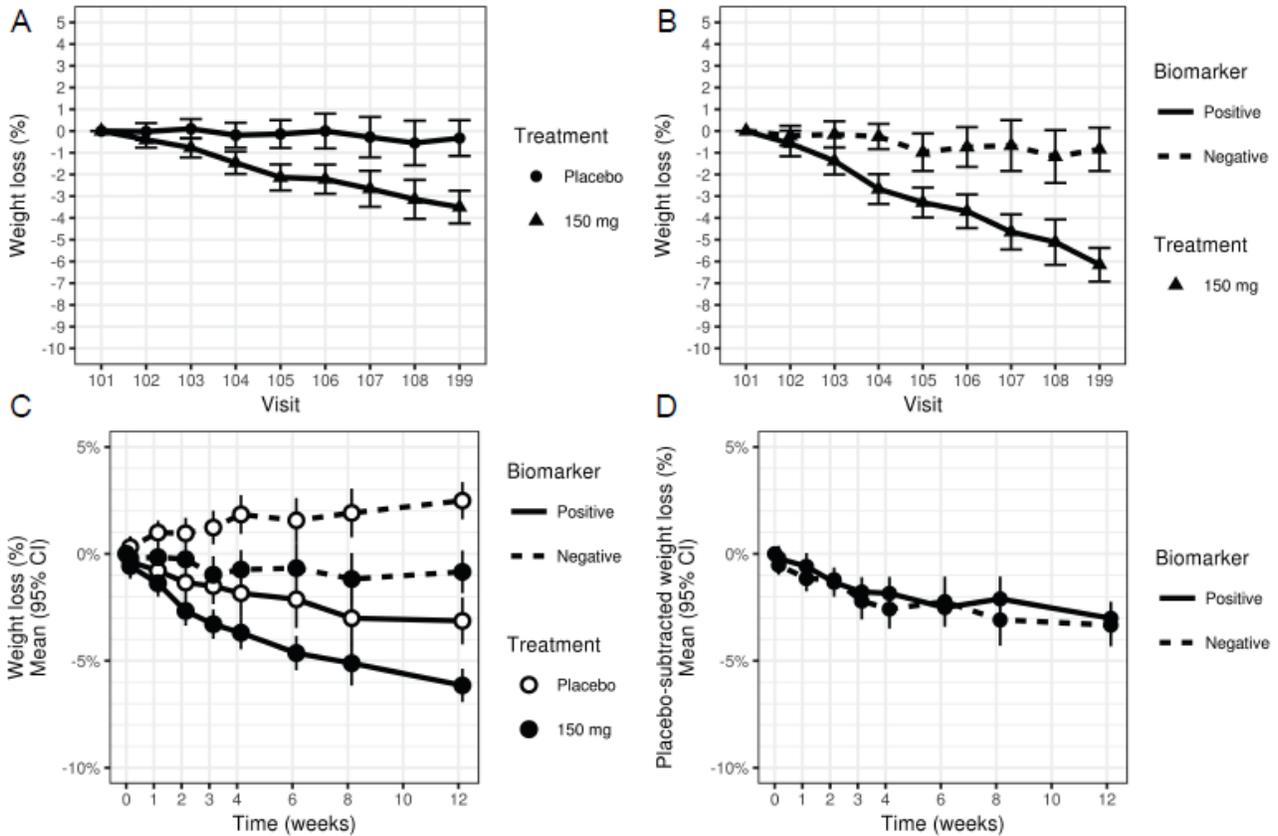
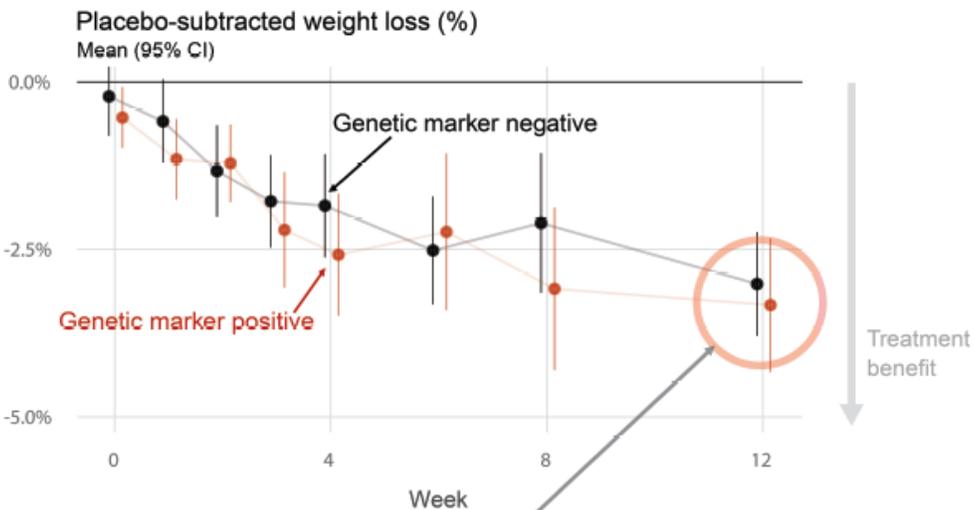

Figure 10: Post-hoc subgroup analysis. A project team saw an overall insufficient treatment effect regarding weight loss (A), but a seemingly stronger benefit in a subgroup of patients identified by a biologically plausible genetic marker (B). Getting the comparison right, however, it becomes clear that the marker, while prognostic for weight loss in general, is not predictive for a treatment effect vs. placebo (C). This is made more obvious by displaying the treatment difference directly (D). Finally, we can craft a messaging slide suited for conveying this message to a group of executives within seconds (E).

**Case study 4: Post-hoc subgroup analysis**

Post-hoc exploratory subgroup analyses are common especially after borderline or failed clinical trials.[57] Often the objective of such analyses is to understand why the study failed, or to identify a subgroup of patients who did show sufficient response to the treatment.[58,59] In this fourth case study we illustrate the challenge of navigating this type of analysis. The objective is to present to decision-makers a recommendation whether or not to proceed with further investigation of a genetic marker that may be predictive of response to treatment.

Figure 10A displays the desired effect, percentage body weight loss from baseline, for an active treatment and a placebo control arm. The primary endpoint is 12 weeks after randomization. Although the active treatment reduced body weight, the average extent of the effect was not considered clinically meaningful. However, the team found a subgroup of patients identified by a biologically plausible genetic marker who appeared to benefit more strongly from the compound, as shown in Figure 10B.

At this point we need to intervene. Already the first plot did not show the data clearly (Law 2), using barely distinguishable plot symbols and huge ticks at the end of the error bars, too many ticks on the y-axis, and – worst offense – visit numbers on the x-axis (equally spaced!). The second plot, in addition, is fundamentally flawed, because it displays the wrong comparison (Law 1, wrong scientific question). The treatment effect in this trial is weight loss under active treatment compared to weight loss under placebo, and this subgroup analysis removes the placebo arm completely.

Let us put the placebo group back in, and at the same time improve the various graphical shortcomings of the plot (using appropriately spaced times on the x-axis, distinguishable plot symbols etc.). We obtain Figure 10C which displays the treatment-by-subgroup interaction over time (and clarifies that the error bars are 95% confidence intervals). Interestingly, we see some weight loss also in the biomarker-positive placebo group, and weight gain in the biomarker-negative placebo group. The treatment effect (active drug vs. placebo) is similar within each biomarker group. In other words, the biomarker appears *prognostic* for weight loss (at least under the conditions of this clinical trial which may include dietary or exercise advice), but not *predictive* for a treatment effect. Figure 10D displays the treatment effect directly, to make it obvious that it is the same irrespective of the genetic marker (Law 3).

To complete our objective, we should present these results and our recommendation about the genetic marker to decision-makers. We need to show the results clearly and make the message obvious. In this context, the message is that the genetic marker is not predictive for a treatment effect; whether it is prognostic for weight loss is less important. Figure 10D supports our message most succinctly. Based on this figure, we finally craft a messaging slide that is suited for conveying this message in a meeting to a group of executives within seconds.

Figure 10E shows this final slide. Guided by Laws 2 and 3, we arrived at this slide by considering the following principles and recommendations.

- *Facilitate comparisons.* We zoom in on the comparison of interest by appropriately restricting the y-axis. A reference line at zero indicates the position of no treatment effect, and a labelled arrow shows the direction of treatment benefit.
- *Reduce cognitive load.* We use direct labelling instead of a legend to identify the two subgroups. We select clearly distinguishable colors that also carry through to the subgroup labels.
- *Adapt to your audience.* We avoid writing the y-axis label vertically, by placing it above the plot and using it as its title. This is to prevent the audience from straining to read information during the meeting. We spell out that this is a treatment comparison (a difference) in terms of weight loss. We also reduce the size of the symbols representing the point estimates, to avoid the audience focusing too much on these and too little on the uncertainty intervals.
- *Remove non-data ink.* We reduce the number of ticks and points displayed on the x- and y-axes. They do not convey useful information for a short presentation, as we do not expect the audience to extract exact values. The connecting lines are pushed to the background by introducing a degree of transparency.

- *Effective redundancy.* We add a caption that spells out what is seen at week 12 (the primary endpoint), and we highlight the corresponding treatment effect estimates by an enclosing circle (a powerful grouping technique to draw attention). Finally, we place the main message directly as the title of the slide (setting the most important word, "**not**", in bold), and as subtitle we explicitly put our recommendation: "Genetic marker does **not** warrant further investigation."

## DISCUSSION

Effective visual communication appears obvious, but creating it requires skill. The purpose of this tutorial was to convey some of this skillset. We have proposed three laws of visual communication for the quantitative scientist:

1. Have a clear purpose
2. Show the data clearly
3. Make the message obvious

The first law is about doing the right thing; the two others are about doing things right. We have distilled a range of more detailed recommendations on the Graphics Principles Cheat Sheet,[50] a concise reference sheet available on https://graphicsprinciples.github.io. Finally, we have discussed four use cases to illustrate the application of these laws and recommendations in practice.

No system is perfect. First, there is simply no tension-free one-size-fits-all approach to visual communication: the given laws and recommendations may occasionally cause conflict. Should we show raw data or summaries? More or less detail? Confidence intervals or not? Individual treatment effects or only their difference? These and other choices depend on the purpose and situation of the communication, the audience, and the author. Importantly, the format of a display or presentation may impose practical constraints (are colors available? how much time will I have? etc.). Good judgement is always due, and sometimes a conscious compromise.

Second, the distinction between the three given laws may be challenged as occasionally blurred:

- We have already mentioned the need to adapt to the audience, which is part of Law 1 (have a clear purpose – with respect to a target audience) as well as Law 3 (make the message obvious – to a specific audience). If your audience does not understand your message, do not repeat it in the same way (and never louder!). Adapt and find other ways to convey it (effective redundancy).
- Another such general principle is: "do no harm." In medical and bio-ethics ("primum non nocere") it is thought to derive from the oath of Hippocrates. But it also applies to graphics and visual communication.[60] For example, do not get carried away by technology (templates, frames, animation etc.) that clouds the data (Law 2) or draws attention away from the message (Law 3; see also Baillie's and Vandemeulebroecke's comment[61] in the discussion of Bowman[62]). The removal of "chart junk" (in Law 2) is another example.
- Bonate[31] mandates a principle which may belong to Law 1 and Law 3: "don't be boring". If you engage your audience through clarity, visual cues, and perhaps humor, it will be easier for them to pay attention and understand your message. Your visual communication will achieve its purpose. However, if you bore your audience with unnecessary detail, an odd visual pace, by clouding your main point or by not even having one, then it will just turn away.
- The Cleveland-McGill effectiveness ranking, introduced in Law 3, serves Law 2 just as much.

Third, we may have missed some special considerations. For example, we did not say much about an appropriate choice of color, considering human color perception, psychological connotations, color-blindness etc. For further information, we refer to Wong,[63] Wilke,[64] and resources such as Munsell[65] and ColorBrewer.[66]

Despite these limitations, we believe that it helps to think of the three proposed laws as major maxims for visual communication for the quantitative scientist. Keeping the Graphics Principles Cheat Sheet "at your fingertips" may provide additional practical support. Finally, our advice is to practice. To quote Tufte:[11] "Graphical competence demands three quite different skills: the substantive, statistical, and artistic." These skills cannot be learned by reading an article. Adopt visualization in every part of your workflow; make it a habit. Think graphically. Use pencil and paper before coding in software.[67,68] Calculate *and* communicate.[68] Test and repeat.

## Acknowledgements

We would like to thank David Carr, Julie Jones, Oliver Sander, Ivan-Toma Vranesic and Andrew Wright.

## Author contributions

All authors have contributed equally to this tutorial.